\documentclass[prb,aps,twocolumn,showpacs,superscriptaddress]{revtex4-1}
\usepackage{hyperref}
\hypersetup{colorlinks, citecolor=blue, filecolor= black, linkcolor= black, urlcolor= blue}
\usepackage{graphicx}
\usepackage{dcolumn}
\usepackage{bm}
\usepackage{float}
\usepackage{amssymb, amsmath}

\begin{document}
	\preprint{Preparing}

	\title{Post-quench dynamics and suppression of thermalization in an open half-filled Hubbard layer}
	
	\author{Igor V. Blinov}
	\email{blinov@phystech.edu}
	\affiliation{
		Moscow Institute of Physics and Technology, 9 Institutskiy lane, Dolgoprudny, Moscow Region 141700, Russia}
	\affiliation{
	Russian  Quantum  Center,  143025  Skolkovo,  Russia}
	
	\author{Pedro Ribeiro}
	\affiliation{
		CeFEMA, Instituto Superior T\'{e}cnico, Universidade de Lisboa Av. Rovisco
Pais, 1049-001 Lisboa, Portugal}
	\affiliation{
		Russian  Quantum  Center,  143025  Skolkovo,  Russia}
		
	\author{A.  N.  Rubtsov}
	\affiliation{
			Russian  Quantum  Center,  143025  Skolkovo,  Russia}
	\affiliation{
	Department  of  Physics,  M.V.  Lomonosov  Moscow  State  University,  119991  Moscow,  Russia}

	\date{\today}
	
	\begin{abstract}
		\noindent

We study the time evolution of an half-filled Hubbard layer coupled to a magnon bath after a quench of the Hubbard interaction. 
Qualitatively different regimes, regarding the asymptotic long time dynamics, are identified and characterized within the mean-field approximation. 
In the absence of the bath, the dynamics of the closed system is similar to that of a quenched BCS condensate. 
Though the presence of the bath introduces an additional relaxation mechanism, our numerical results and analytical arguments show that equilibration with the bath is not necessarily attained within approximations used. Instead, non-equilibrium states, similar to the ones observed in the closed system, can emerge at long times as a consequence of the competition between intrinsic relaxation mechanism (Landau damping, for example), and the bath-induced dissipation.
	\end{abstract}

	\maketitle
	\newpage
	
	
\section{Introduction}
Most thermodynamic systems, if taken away from equilibrium, evolve back to an equilibrium state.  The initial stage of the equilibration process often includes energy transfer from macroscopic collective excitations to the individual, microscopic degrees of freedom. The overall equilibration dynamics is thus governed by the coupling between the collective and individual modes. 

Modern quantum technologies aim to store information via controlled excitation of collective states in engineered  solid structures, such as various types of superconducting qubits. 
In order to improve quantum coherence, significant attention, both at experimental and theoretical levels, has been paid to the optimization of the decoupling between collective modes and the rest of the degrees of freedom. 
In certain cases, undamped collective excitations completely decoupled from the microscopic degrees of freedom were predicted theoretically. Such decoupling arises due to existence of conservation laws present in some particular system\cite{polkovnikov2011colloquium}. Although the optimal degree of decoupling is model-specific and the excitation lifetime is, in practice, always finite, one expects those systems to show considerable improvements in experimentally observed decoherence times.

The energy transfer from the collective to individual degrees of freedom can, in many cases, be studied using the concept of Landau damping, which does not require a detailed knowledge about the decoherence process. 
 Landau damping appears in a collisionless models and its only precondition is the causality principle. 
First formulated for Langmuir waves in a collisionless electron plasma \cite{landau1946vibrations},  Landau damping is nowadays known to be a generic feature of the mean-field perturbative description of collective excitation. In particular, it appears in the BCS description of superconductors \cite{gor1968generalization, volkov1973collisionless}. In this case, the collective mode is associated with deformations of the superconducting order parameter $\Delta$ and the individual degrees of freedom are Cooper pairs. 

Sufficiently far away from equilibrium, regimes beyond the Landau-damping scenario may arise, such as the ones found in recent studies\cite{levitov, altshuler} of the BCS-model. Here, a perturbation of the initial ground state is realized as an abrupt change (quench) of the BCS coupling parameter $g$ from its initial value $g_i$ to a final one $g_f$. 
For a small perturbation $g_i\approx g_f$  the dynamics can be well described by a Landau-damping scenario \cite{volkov1973collisionless}. 
For $g_i \ll g_f$, as in the case of the quench from normal metal to BCS \cite{barankov2004collective}, a synchonization between Cooper-pairs through the collective mode yields to persistent oscillations of the order parameter (phase-locked regime). 

In the opposite case, $g_i \gg g_f$, the gap vanishes exponentially fast\cite{levitov} (overdamped regime), since the system is effectively heated above the superconducting transition temperature.

The interplay between microscopic and collective modes has been studied in other setups \cite{bulgac2009large, foster2013quantum, foster2014quench, Peronaci2015, chichi, Gao2014, dzero2015amplitude}, 
including the non-equilibrium dynamics of half-filled Hubbard model supporting an antiferromagnetic collective modes. 
Within the Gutzwiller approach, the after-quench dynamics of this model was shown \cite{sandri} to be similar to that of the BCS model featuring all three regimes. However, because of electron-electron collisions, oscillations were found to be weakly damped in the phase-locked regime. The transition between the anti-ferromagnetic and paramagnetic states was also explored \cite{tsuji} within DMFT.

The examples given below refer to close systems. Under which conditions the asymptotic long time state equilibrates has recently been an active topic of research  \cite{Srednicki1994, Deutsch1991, Rigol2008}. Clearly, some of the above regimes cannot be considered as equilibrium states. Nonetheless, if equilibrium is attained, the extensive injection of energy implies that properties of the effective equilibrium state correspond to those of a finite temperature Gibbs-like ensemble\cite{Rigol2007, Ilievski2015}.

The presence of a weakly-coupled zero-temperature reservoir is expected to radically change the physical picture\cite{Amaricci2012, Mazza2015}: if energy is dissipated to the bath, the system's degrees of freedom should acquire properties of the post-quench zero-temperature state. However, a competition between the system's own dissipative  processes and those of the bath may allow for other scenarios. 
Understanding the robustness of different dynamical regimes to the presence of an environment is a natural question as in realistic experimental situations some degree of environmental coupling is expected.
The competition between dissipative effects induced by the microscopic degrees of freedom of the system or of the bath can help to shed light in the collective mode dynamics observed in recent pump-probe-like experiments. 

In this work we study the fate of non-equilibrium regimes, found in the post-quench dynamics of closed system, in the presence of a bath. 
In particular, we consider interaction quenches in the half-filled Hubbard model on a 2d square lattice, coupled at each site to a magnetic Ohmic bath. The bath degrees of freedom consist of a collection of vector bosons that couple isotropically to the local magnetization and are taken to be independent on each lattice site. The system models an antiferromagnetically ordered 2d layer in the presence of a magnon environment. 
Such spatially independent environment models a superparamagnetic bath medium present for example in disordered nanomagnets. In addition, though for an homogeneous ordered substrate spatial correlations of the bath modes may become important, our results are still relevant if the order of low-lying bath modes is incommensurate with that of the Hubbard layer. In this case, although the spatially coherent states still couple to the magnetic modes of the Hubbard layer their ordering is not transferred to the layer.


The paper is organized as follows: Sec. II introduces the model, Sec. III describes the dynamics in the absence of any environmental coupling and identifies the different dynamical regimes in correspondence with the one in the BCS model,  section IV presents the bulk of our work identifying the different dynamic regimes in the presence of the bath. A discussion and conclusions are given in Sec. V. The appendix is devoted to study the specificities of the overdamped regime in the 2d square lattice.

\section{Model}
To study the post-quench dynamics we consider a joint Hamiltonian of the antiferromagnetically ordered layer coupled to a magnon bath given by 
\begin{equation}
H=H_{\text{Hub}}+H_{\text{Bath}}+H_{\text{C}},
\end{equation}
where 
\begin{equation}\label{model:hubbard}
H_{\text{Hub}}=
J\sum_{\left<\boldsymbol{r},\boldsymbol{r}'\right>}
c^{\dagger}_{\sigma \boldsymbol{r}}c_{\sigma \boldsymbol{r}}
+
U \sum_{\boldsymbol{r} }
c^{\dagger}_{\uparrow,\boldsymbol{r}}c_{\uparrow \boldsymbol{r}}c^{\dagger}_{\downarrow \boldsymbol{r}}c_{\downarrow \boldsymbol{r}}
\end{equation}
is the two-dimensional Hubbard Hamiltonian describing the electronic system.  $c^{\dagger}_{\sigma \boldsymbol{r}}$ and $c_{\sigma \boldsymbol{r}}$ are, respectively, the fermion creation and annihilation operators of an electron on site $\boldsymbol{r}$ with spin $\sigma$. 
The coupling to the bosonic bath is given by:
\begin{equation}\label{model:coupling}
H_{\text{C}}=
g
\sum_{\boldsymbol{r}} 
\boldsymbol{S}_{\boldsymbol{r}}
\cdot \left[ \int_q \left(
\boldsymbol{b}^\dagger_{q \boldsymbol{r}}+\boldsymbol{b}_{q \boldsymbol{r}} \right)
\right],
\end{equation}
where $S_{\boldsymbol{r}}^i=\frac{1}{2}c^\dagger_{\alpha \boldsymbol{r}} \tau_{\alpha\beta}^i c_{\beta \boldsymbol{r}} $ is the spin operator of the electrons at the $\boldsymbol{r}$-th site and $\tau_{\alpha\beta}^{i=x,y,z}$ denote the Pauli matrices. $b^{i \dagger}_{q \boldsymbol{r}}$, ${b}^i_{q \boldsymbol{r}}$ are creation and annihilation operators of the vector bosons. The magnon environment is  assumed to be spacially incoherent, therefore the summation over bosonic momentum index $q$ is performed independently for each site index $ \boldsymbol{r}$.  
The Hamiltonian of the bath is 
\begin{equation}\label{model:bosons}
H_{\text{Bath}}=
\sum_{\boldsymbol{r}} \int_q
\Omega_q
\boldsymbol{b}^\dagger_{q \boldsymbol{r}} \cdot
\boldsymbol{b}_{q \boldsymbol{r}}.
\end{equation}
The two dimensional antiferromagnet magnon bath has an Ohmic density of states of the form $\rho(\epsilon)= \int_q \delta(\epsilon-\Omega_q)=\frac{\epsilon}{C^2} \, e^{-\frac{\epsilon}{\Lambda}}$, where $\Lambda$ is a high-energy cutoff and $C$ is a constant with dimension of energy. In the following we set $C=1$ and measure all other energies in units of $C$. In all numerical result we set $\Lambda=20$, $J=2$. No qualitative changes arise by changing the cutoff $\Lambda$ as long as it is taken to be much larger than all other energies scales. 

We study the dynamics of the magnetization after an interaction quench where the value of $U$ is switched from $U=U_i$ 
 to $U=U_f$ at $t=0$.  We treat the system within a mean-field approximation assuming the spin fluctuations are small as compared with the average magnetization. For all the analyzed cases the initial state is the ground state  of the system for $U=U_i$ obtained within the mean-field approach.

\section{Closed system}
In this section we study quenches of the closed system, i.e. when the coupling to the bath, $g$, is set to zero. First, we derive the set of mean-field equations governing the post-quench dynamics. We then analyze the long time asymptotic dynamics, establishing the parallels and differences to previous studies of the BCS model. 

For the isolated electronic system described by Eq.\eqref{model:hubbard} the mean-field approximation is valid for time scales lesser than $\tau_q\sim E_F/\Delta^2$,  after which, interactions between quasiparticles can no longer be neglected \cite{barankov2004collective}. 
The mean-field Hamiltonian can be obtained from Eq.\eqref{model:hubbard} by neglecting second order magnetic fluctuation terms $(\boldsymbol S_{\boldsymbol{r}}-\left< \boldsymbol S_{\boldsymbol{r}}\right>)^2$ and assuming a spin ordered state $\left<\boldsymbol S_{\boldsymbol{r}}\right>= \boldsymbol M \cos(\boldsymbol{Q}\cdot\boldsymbol{r})$ with ordering wavevector $\boldsymbol{Q}$ \cite{fradkin}. 

A systematic approach for the construction of a mean-field approximation can be obtained within the functional integral formalism and amounts to a decoupling of fermionic fields using the Hubbard-Stratonovich transformation \cite{altland2010condensed}. The latter step, while mathematically exact,  induces an ambiguity (sometimes alluded to as Fierz ambiguity \cite{fierzAmbiguity}) related with the choice of the decoupling channel that conditions further approximations. In this work, we deal exclusively with an half-filled Hubbard layer known to have an magnetic instability towards the formation of an antiferromagnetic state. Therefore, on physical grounds, we have have chosen a decoupling in the magnetic exchange channel as this gives the leading contribution to free energy. 
We thus consider an antiferromagnetic state, i.e. $\boldsymbol{Q}=\{\pi,\pi\}$, magnetised along the $z$-axes, $\boldsymbol M = M \boldsymbol e_z$, corresponding to a mean-field Hamiltonian of the form
\begin{equation}\label{AFM-hamiltonian}
	H_\text{MF}=
	\int_k  \epsilon_{\boldsymbol k} c^{\dagger}_{\sigma \boldsymbol k}c_{\sigma \boldsymbol k}
	+\frac{2U}{3}M^2
	-\frac{4U}{3}S^z_{\boldsymbol Q} M
\end{equation}
with $\boldsymbol k$ labelling the two-dimensional momentum and $\int_k = \int \frac{d^2 k}{(2\pi)^2}$ the Brillouin-zone integration.  $\epsilon_{\boldsymbol k}  = 2J (\cos k_x+ \cos k_y)$ is the dispersion relation and  $S_{\boldsymbol{Q}}^z = \frac{1}{2}\int_{k}
c^{\dagger}_{\sigma \boldsymbol k}  \tau^z_{\sigma \sigma'} c_{\sigma' \boldsymbol k + \boldsymbol Q}$ the staggered spin-operator in the $z$ direction.

The dynamics can most easily  be  described in terms of pseudo-spins, that are similar to Anderson representation \cite{anderson1958random}:
\begin{align}
\label{tau_x}
\hat \tau^x_{\sigma \boldsymbol k}=
\frac{1}{2}
(c_{\sigma \boldsymbol k}^\dagger c_{\sigma \boldsymbol k + \boldsymbol Q}+
c_{\sigma \boldsymbol k + \boldsymbol Q}^\dagger c_{\sigma \boldsymbol k}
)\\
\label{tau_y}
\hat \tau^y_{\sigma \boldsymbol k}=
\frac{i}{2}
(
c_{\sigma \boldsymbol k}^\dagger c_{\sigma \boldsymbol k+ \boldsymbol Q} -
c_{\sigma \boldsymbol k+ \boldsymbol Q}^\dagger c_{\sigma \boldsymbol k}
)
\\
\label{tau_z}
\hat \tau^z_{\sigma \boldsymbol k}=
\frac{1}{2}
(
c_{\sigma \boldsymbol k}^\dagger c_{\sigma \boldsymbol k}-
c_{\sigma \boldsymbol k + \boldsymbol Q}^\dagger c_{\sigma \boldsymbol k+\boldsymbol Q}. 
)
\end{align}
defined for each spin projection $\sigma$. 
Assuming that the initial state respects the symmetries of $H_\text{MF}$, the expectation values for the two spin projections are simply related by
\begin{align}
\left<\hat \tau^x_{\uparrow \boldsymbol k }\right>
=
-\left<\hat \tau^x_{\downarrow \boldsymbol k }\right>,
\\
\left<\hat \tau^y_{\uparrow \boldsymbol k }\right>
=
-\left< \hat \tau^y_{\downarrow \boldsymbol k }\right>,
\\
\left< \hat \tau^z_{\uparrow \boldsymbol k }\right>
=
\left<\hat \tau^z_{\downarrow \boldsymbol k }\right>.
\end{align}
Therefore, in the following we set  $\hat \tau^{\alpha}_{\boldsymbol k} = \hat \tau^{\alpha}_{\uparrow \boldsymbol k}$ in order to simplify notation.
In terms of the pseudospin variables $\tau$ the equations of motion can be written in a closed form
\begin{equation}\label{bare-dynamics}
\frac{d}{dt}\left<\hat{ \boldsymbol{  \tau }} _{\boldsymbol k} (t) \right>=2\boldsymbol{B}_{\boldsymbol k}(t)\times\left< \hat{\boldsymbol{\tau}}_{\boldsymbol k} (t)\right>
\end{equation}
where 
$
\boldsymbol{B}_{\boldsymbol k}(t) =\{h_c(t),0,\epsilon_{\boldsymbol k}\},
$
with $h_c(t) =-4U(t) M(t)/3$ and the self-consistent condition 
\begin{equation}\label{self-consistence}
M(t)=\int_k  \left< \hat \tau^x_{\boldsymbol k}(t) \right>.
\end{equation} 
For the quench protocol studied here $U(t>0)=U_f$.
The initial conditions, obtained by starting from the ground-state of $H_{\text{MF}}$ with $U(t=0)=U_i$, are given by
\begin{equation}\label{initials}
\left< \hat{ \boldsymbol{\tau}}_{\boldsymbol k}  (t=0) \right>  = - \frac{\boldsymbol{B}_{\boldsymbol k}(t=0)}{2 \sqrt{\epsilon_{\boldsymbol{k}}^2+h_c(t=0)^2}}. 
\end{equation}
This result can be derived by minimizing the mean-field energy, that in terms of the pseudospin $\tau$ is given by 
$E_{\text{MF}}=\left<  H_{\text{MF}}\right> =2\int_k \boldsymbol{B}_{\boldsymbol{k}}.\left<\hat{\boldsymbol{\tau}}_{\boldsymbol{k}}\right>$
with respect to the order paramter $M(t=0)$. 

The mean field dynamics obtained by this procedure is closely related to that of the BCS-model \cite{levitov,tsuji, sandri}. In fact, the equations of motion \eqref{bare-dynamics}  and those of the BCS Hamiltonian can be mapped to each-other by a suitable identification of physical quantities.  
The main difference between the two models comes from the dispersion relation $\epsilon_{\boldsymbol{k}}$ that in the BCS-model is usually taken to be that of a free-electron gas, yielding in 2d to a constant density of states within the Debay window. 
Here, the fact that $\epsilon_{\boldsymbol{k}}$ admits the nesting wave-vector $\boldsymbol Q$ at half filling is crucial for the establishment of the anti-ferromagnetic instability and has therefore to be explicitely taken into account. 

In the reminder of this section we study the different dynamical regimes of the asymptotically large time dynamics of the post-quench evolution governed by the equations \eqref{bare-dynamics}. The different regimes  are similar to those of the BCS model \cite{tsuji}, crucial differences arise nonetheless in the approach to the long-time limit due to the particular features of the dispersion relation.

\begin{figure}[htb!]
	\centerline{\includegraphics[width=8.5cm]{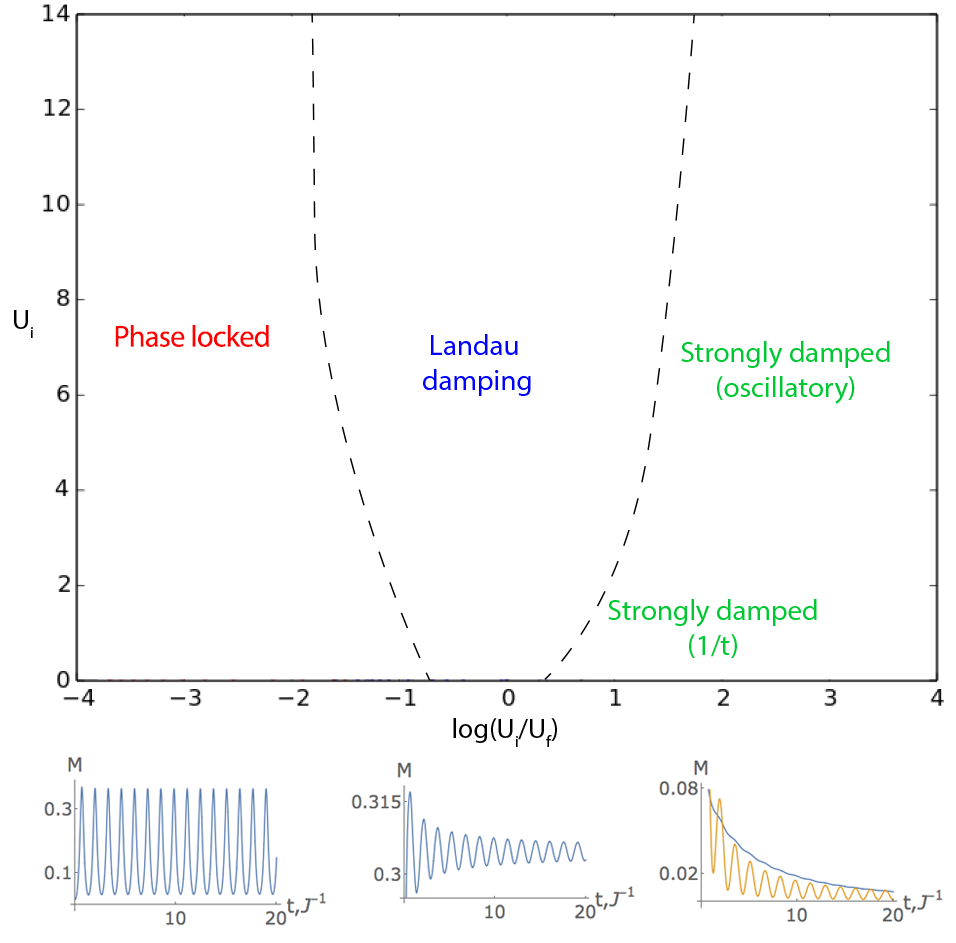}} 
	\caption{Upper panel: Sketch of the phase diagram as a function of $U_f$ and $U_i$. 
	Lower pannel: Examples of the different dynamical regimes, from left to right: phase-locked - $U_i=0.8$, $U_f=12$; Landau-damping - $U_i=4$, $U_f=5$; overdamped - $U_i=3$, $U_f=0.5$. \label{phase_diagram_close} 
	}
\end{figure}
Fig. (\ref{phase_diagram_close})-upper panel shows the phase diagram in the $U_f-U_i$ parameter space.  
As in the BCS case there are three different regimes, shown in the Fig. (\ref{phase_diagram_close})-lower panel:
\\\\
- The phase-locked regime, arising for $U_i/U_f\ll 1$, is characterised by non-vanishing oscillations of the order parameter. This behavior is similar to the one described in \cite{barankov2004collective}: the collective mode synchronizes the different momentum pseudo-spin precessions. 
\\\\
- The Landau-damping regime, for $U_i/U_f\approx1$, where the order parameter attains a non-vanishing constant value. 
Here, oscillations decay as $\propto\frac{1}{\sqrt{t}}$ as in the BCS case \cite{yuzbashyan2006dynamical}. 
The mechanism behind this kind of damping is similar to the one firstly found in plasma \cite{landau1946vibrations}: as in BCS case \cite{gor1968generalization, levitov}, a collective mode interacts with quasiparticles with energies around $2\Delta$, where $\Delta$ is an antiferromagnetic gap.  \\\\
- The overdamped regime, when $U_i/U_f\gg 1$, where the order parameter vanishes at large times. 
As in the BCS case \cite{yuzbashyan2006dynamical} the order parameter drops to zero. However, instead of the exponential decay observed for BCS, the decay is algebraic in $1/t$ and a crossover is observed as a function of $U_i$ from damped-oscillatory to purely damped behaviour in the dynamics of $M(t)$. 
This behaviour, overlooked in similar setups \cite{sandri,tsuji}, is due to the non-analyticities of the density of states in two dimensions: a logarithmic divergence near the Fermi surface, and a sharp cutoff at the band edges.  A detailed analysis of the crossover is given in Appendix \ref{Crossover}.

\section{Open system}
We now address the changes in the dynamics of the system in the presence of the environment. In the following we generalize the equations of motion to account for the magnetic bath and analyze the different dynamical regimes. 

Following the same steps as before, the mean-field Hamiltonian is given by:
\begin{multline}\label{coupled-hamiltonian}
H_{\text{MF}}= \int_k 
\epsilon_{\boldsymbol k} c^{\dagger}_{\sigma \boldsymbol k} c_{\sigma \boldsymbol k}
+\frac{2U}{3}M^2-\frac{4U}{3}S_{\boldsymbol Q}^z M \\
+ g M \int_q ({b}^{z\dagger}_q+b_q^z)
+\int_q \Omega_q {b}^{z\dagger}_q b_q^z.
\end{multline}
In addition to the pseudo-spin degrees of freedom the dynamics of the bosonic fields also has to be considered. 
At the mean-field level this can be done by explicitly solving the equation of motion for the bosonic fields  
$\frac{d}{dt}b_q^z=i[H_{\text{MF} },b_q^z]$. Similarly to the case of the closed system, the initial state is taken to be the ground state of the whole system.
Substituting in the equations of motion of pseudo-spins we get a closed system of equations that have the same form (\ref{bare-dynamics}) as in the isolated case but with a different "pseudo magnetic field" $\boldsymbol{B_{\boldsymbol k}}(t)=\{h_d(t),0,\epsilon_k\}$ where 
\begin{multline}\label{open:pseudofield}
h_d(t)=
-\frac{2U(t)M(t)}{3}
-g^2\frac{\Lambda  M(0)}{1+\Lambda^2t^2} \\
-2g^2\int^t_0d\tau\frac{\Lambda^3M(\tau)(t-\tau)}{[1+\Lambda^2(t-\tau)^2]^2}.
\end{multline}
 As before, $M(t)$ respects the self-consistency condition (\ref{self-consistence}) and $M(0)$ is the initial value of the staggered magnetization.

Before studying the effects of the environment in different dynamic regimes let us analyze the stationary solutions $M(t)\equiv M$ in the presence of the environment.
In this case $h_d$ simplifies to 
\begin{equation}\label{open:pseudofield:stat}
h_d =-2\left(\frac{2U}{3}+g^2\Lambda \right)  M.
\end{equation}
This stationary condition is equivalent to that of the close system with a renormalisation of the value of the coupling $U\rightarrow U_{\text R}=U+\frac{3g^2\Lambda}{2}$.
Therefore, the only effect of environment on equilibrium properties of electronic subsystem is a renormalization of the coupling constant.
Since the renormalization of $U$ is always positive, the presence of the environment always enhances the antiferromagnetic order. 

It is worth noting that in case the system approaches such stationary solution (not necessary an equilibrium one) the equations of motion of the open system reduce to those of the closed one with a renormalized $U$.
This can be most easily shown by introducing a time scale $T_{\text{stat}}$ after which $M(t)$ is close to the stationary value $M_{\text{stat}}$.
For times $1/\Lambda \ll T_{\text{stat}} \ll t$, up to terms of order $t/T_{\text{stat}} $ and $1/(\Lambda t)^2$, one has:
\begin{multline}\label{open:pseudofield:eqiulibrium}
h_d (t)
\approx 
h_{\text{stat}}
-\frac{2g^2 \Lambda [M(0)-M_{\text{stat}}]}{(\Lambda t)^2} \\
-\frac{4g^2}{(\Lambda t)^3}\int^{T_{\text{stat}}}_0d\tau M(\tau),
\end{multline}
which when $t\rightarrow\infty$  tends to $h_{\text{stat}} = -\frac{4U_f M_{\text{stat}}}{3}
-2g^2\Lambda M_{{\text{stat}}}$.
Thus, for sufficiently large times, in the approach to $M_{{\text{stat}}}$ the individual degrees of freedom $\tau_k^x$ are governed by the renormalized electronic dynamics. Consequently, we may conclude, that whenever configuration with stationary $M_{\text{stat}}$ (not necessarily equilbrium) has been reached, environment's role is reduced to renormalization of $U$. This argument is essential for undersanding absence of thermalization in overdamped and damped regimes which will be described below.

In the following it is important to distinguish between two kinds of stationary solution:  equilibrium states where $M_{\text{stat}}$ minimizes the mean-field energy and yields no dynamics to the pseudo-spins $d \left< \hat{\boldsymbol \tau }_{\boldsymbol k }  \right> /dt= 0$; and non-equilibrium states where $d \left< \hat{\boldsymbol \tau }_{\boldsymbol k }  \right>/dt \neq 0$. 
In the case of a closed system conservation of energy implies that only non-equilibrium stationary states can be attained as is the case of the final state of the Landau damped regime. 
In the presence of a bath, even if the total energy is still conserved, a change of energy of the system can be absorbed by the bath with no macroscopic changes in any intensive bath observable. It is naively expectable that, by absorbing the excess energy, the environment renders the system observables to their equilibrium values. 
Nonetheless, as shown below, both equilibrium and non-equilibrium solutions may arise for the open system. 

\begin{figure}[htb!]
	\centerline{\includegraphics[width=0.5\textwidth]{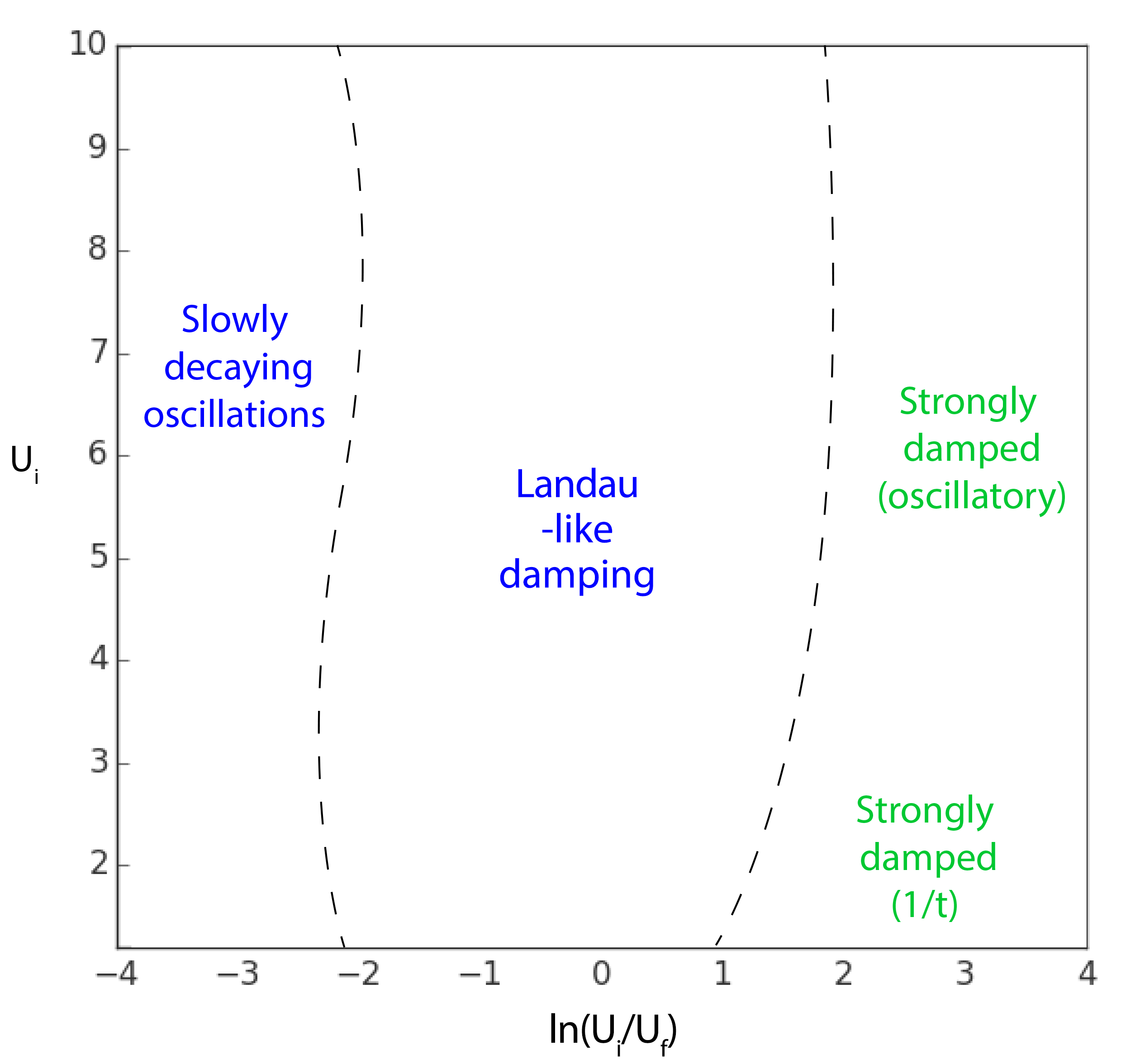}} 
	\caption{
	Sketch of the phase diagram for open system for $g=0.25$. 
	With increasing of $g$ the damped regime expands, pushing its boundaries in both directions.
	This arises due to the renormalization of $U$ (see text) that decreases the effective amplitude 
	of the quench. 
	}
	\label{fig:open:phase-diagram}
\end{figure}

Fig. \ref{fig:open:phase-diagram} shows a sketch of the phase diagram of the open system for different values of the coupling $g$. Approximate boundaries between phases were estimated using $N=150$. In particular, the boundary between the regime with slowly decaying oscillations and the Landau-like damped case was estimated by plotting the order parameter $\Delta M(T_N) = [M(T_N) - M_{\text{eq}}]/M_{\text{eq}}$, where $T_N\propto N$ is the largest time for which the evolution does not depict any finite size effects (see sections IV and Fig. \ref{fig:open:lockphase:panel} for details). With the present numerical data, one cannot determine boundaries precisely, therefore the sketch in Fig. \ref{fig:open:phase-diagram} provides only a qualitative understanding of their mutual arrangement. The non-monotonic behaviour of the left boundary can be an artefact of the method.
The three phases found are reminiscent of those described for the closed system. In the following sections we present our numerical results obtained by solving the equations of motion and give analytical arguments in order to characterise the nature of each phase. 

The equation of motion (\ref{bare-dynamics}), with the memory kernel defined in Eq.\eqref{open:pseudofield}, were solved numerically using a 4th-order Runge-Kutta method. The integral in Eq.\eqref{open:pseudofield} was calculated at each step by employing Simpson's rule.  Calculations were performed on a discrete momentum-grid corresponding to a finite system with periodic boundary conditions and linear size $N$. Accordingly,  $k$-space integrals were substituted by discrete sums: $\int_k \to \frac{1}{N^2} \sum_k$. All the numerics were done using $J=2$.

\subsection{Damped regime} \label{sec:damped_regime}

\begin{figure}[htb!]
	\centerline{\includegraphics[width=0.5\textwidth]{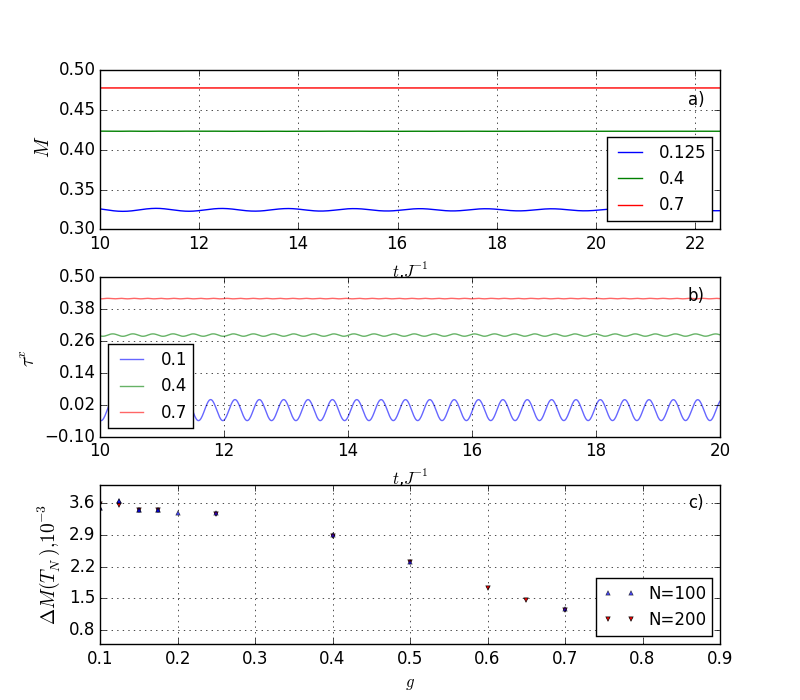}} 
	\setlength{\abovecaptionskip}{1pt plus 1pt minus 2pt}
	\caption{Impact of the bath on the post-quench dynamics. Damped regime decribed in sec. \ref{sec:damped_regime}, for $U_i=4.5$ to $U_f=5$, $\Lambda=20$. 
	a) Time evolution of staggered magnetization for several values of $g$ and $N=200$.
	b)  Time evolution of individual mode $\tau^x_k$ for several values of $g$ and $N=200$
	computed for a generic value of the momentum $k=\{ 2\pi/(N-1),\pi/(N-1) \}$. 
	c) Dimensionless parameter $\Delta M(T_N)$ as a function of $g$ for two values of $N$. The finite size times were taken to be $T_{N=100}=23$, $T_{N=200}=45$. 
\label{fig:open:landau:panel}
	}
\end{figure}
As argued before, in the mean-field description of the open system the presence of the dissipative bath only seems to qualitatively affect the evolution as long as $M(t)$ is time dependent; for $M(t)=M$ its effect simply amounts to a renormalization of the interaction constant. 
This helps to understand  $U_f \simeq U_i$ quenches, corresponding to the Landau-damping regime in the closed system. 
Fig. \ref{fig:open:landau:panel}-a)  shows the time evolution of the order parameter for different values of the environment coupling $g$. As in the closed case one observes a decay of the persistent oscillations and the establishment of an asymptotic stationary state that differs from the equilibrium one. 
Fig. \ref{fig:open:lockphase:panel}-b) further corroborates that $M_{\text{stat}}\neq M_{\text{eq}}$ as the dynamics of the pseudo-spins seems to be non-trivial $d \left< \hat{\boldsymbol \tau }_{\boldsymbol k }  \right> /dt \neq 0$ in the long lime limit. 

Fig. \ref{fig:open:landau:panel}-c) shows the rescaled deviation from equilibrium of the order parameter $\Delta M(T_N) = [M(T_N) - M_{\text{eq}}]/M_{\text{eq}}$ as a function of $g$ for different system sizes. 
$T_N\propto N$ is the largest time for which the evolution does not depict any finite size effects. Defined in this way $\Delta M_\infty=\lim_{N\to\infty}\Delta M(T_N)$ vanishes in the equilibrated phase and is non-zero for non-equilibrium stationary solutions. We observe that $\Delta M(T_N)$ appears to be converged to $\Delta M_\infty$ for the considered sizes. The decreasing of $\Delta M_\infty$ with the coupling to the bath is due to the fact that for larger $g$ both the initial and the final values of the renormalized $U$ increase with $g^2$, therefore the relative quench magnitude decreases and thus, in the large $g$ limit the quenched system is asymptotically  close to the equilibrium one.

\begin{figure}[htb!]
	\centerline{\includegraphics[width=0.4\textwidth]{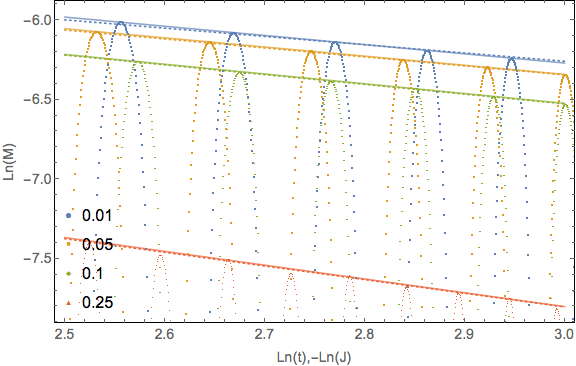}} 
	\setlength{\abovecaptionskip}{1pt plus 1pt minus 2pt}
	\caption{Influence of bath on exponents $\nu_N(g)$ in the damped regime ($U_i=4.5$ to $U_f=5$, $\Lambda=20$) for different values of coupling $g$ and different system sizes (solid line for N=$150$ and dotted for $N=100$, coincide for all $g$-s). 	Exponents were computed by fitting maximums of staggered magnetization to linear function $-\nu_N(g) \log(t)+a$ (solid line). 
\label{fig:open:landau:exponent}
	}
\end{figure}

In order to understand this behaviour we proceeded as in the closed case and consider the quench to be a small perturbation  $ \delta M/M(\infty) \ll1$,  with  $\delta M=M(\infty)-M_{\text{eq}}$, where $M_{\text{eq}}$ is the equilibrium value of the magnetization at $U=U_f$.
The solution of the equations of motion \eqref{bare-dynamics} is assumed to be of the form
 $\left< \hat{\boldsymbol{\tau_{\boldsymbol k}}} (t) \right> =
 \left< \hat{\boldsymbol{\tau_{\boldsymbol k}}} \right>_{\text{eq}} +\boldsymbol{s_{\boldsymbol k}}(t)$, 
 and  $M(t)=M_{\text{eq}}-\delta(t)$, where $ \left< \hat{\boldsymbol{\tau_{\boldsymbol k}}} \right>_{\text{eq}} $ is equilibrium value of pseudo-spin for $U=U_f$.
Expanding Eq. \eqref{bare-dynamics}  to first order in for $s^x_{\boldsymbol k}$, $\delta (t)$ and $\delta M$ one obtains:
\begin{equation}
\frac{d}{dt}\boldsymbol{s_k}(t)
\approx
2
\left(
    \begin{array}{c}
     b_x(t) \\
      0\\
      \epsilon_k
    \end{array}
  \right)
\times\boldsymbol{s_{\boldsymbol k}}(t)
\end{equation}
with $b_x(t) = -\frac{4UM_{\text{eq}}}{3}+2g^2\frac{\Lambda  \delta M}{1+\Lambda^2t^2}-2g^2\Lambda M_{\text{eq}}$. 
Further simplifying the equation  by assuming $t\rightarrow\infty$ one gets explicitely 
\begin{equation}\label{open:linear:system:big-t}
\begin{array}{lll} 
\frac{d}{dt}s^x_{\boldsymbol k}(t)&\approx&-2\epsilon_{\boldsymbol k} s_{\boldsymbol k}^y
\\ 
\frac{d}{dt}s^y_{\boldsymbol k}(t)&\approx&2\epsilon_{\boldsymbol k} s_{\boldsymbol k}^x+2 b_x s_{\boldsymbol k}^z
\\
 \frac{d}{dt}s^z_{\boldsymbol k}(t)&\approx&-2 b_x s_{\boldsymbol k}^y
\end{array} 
\end{equation}
with $b_x = \lim_{t \to \infty} b_x(t)= -\frac{4UM_{\text{eq}}}{3}-2g^2\Lambda M_{\text{eq}} $. 
The solution for $s^x$ is thus of the form:
\begin{equation}\label{open:linear:sx:solution:big-big-t}
s^x_{\boldsymbol k}(t)
\approx
C_{\boldsymbol k} \frac{
\epsilon_{\boldsymbol k}  \cos\left[
2\sqrt{\epsilon_{\boldsymbol k}^2
+\left(\frac{4U_{\text R} M_{\text{eq}} }{3}\right)^2}t \right] }{\sqrt{\epsilon_{\boldsymbol k}^2
+\left(\frac{4U_{\text R} M_{\text{eq}} }{3}\right)^2}  
}
\end{equation}
where the constants  $C_{\boldsymbol k}$ are determined by the initial condition and the previous evolution of the system for times smaller  than $t\simeq \Lambda^{-1}$. 
The form of the Eq. \eqref{open:linear:sx:solution:big-big-t} is the same as the one for the closed system \cite{yuzbashyan2006dynamical} with $U_f$ substituted by $U_{\text{R}}$. Besides this renormalization factor, the only impact of the bath is accounted in the coefficients $C_{\boldsymbol k}$. 
Thus,  in this regime, the dynamics of the individual degrees of freedom of the open system are qualitatively similar to that of the closed one. 
Nonetheless the dependence of $C_{\boldsymbol k}$ on the bath makes the exponent $\nu$, governing the approach to the asymptotic value $M(t) \simeq M(\infty) + O(t^{-\nu}) $,  different from the Landau-damping result $\nu=1/2$. 
Fig. (\ref{fig:open:landau:exponent}) shows a log-log plot of the staggered magnetization as a function of time. In order 
to estimate $\nu$ we fit the local maxima of to the function $-\nu_N(g)\log(t)+a$.  
The exponent $\nu$ is found to have a substantial dependence on $g$: it varies from $\nu_{200}(g=0.0)\approx 0.5$ to  $\nu_{200}(g=0.25)\approx 0.9$ smoothly. Finite-size effects were found to be negligible for $N=150$ and $N=200$.

\subsection{Equilibrating regime}
\begin{figure}[htb!]
	\centerline{\includegraphics[width=0.5\textwidth]{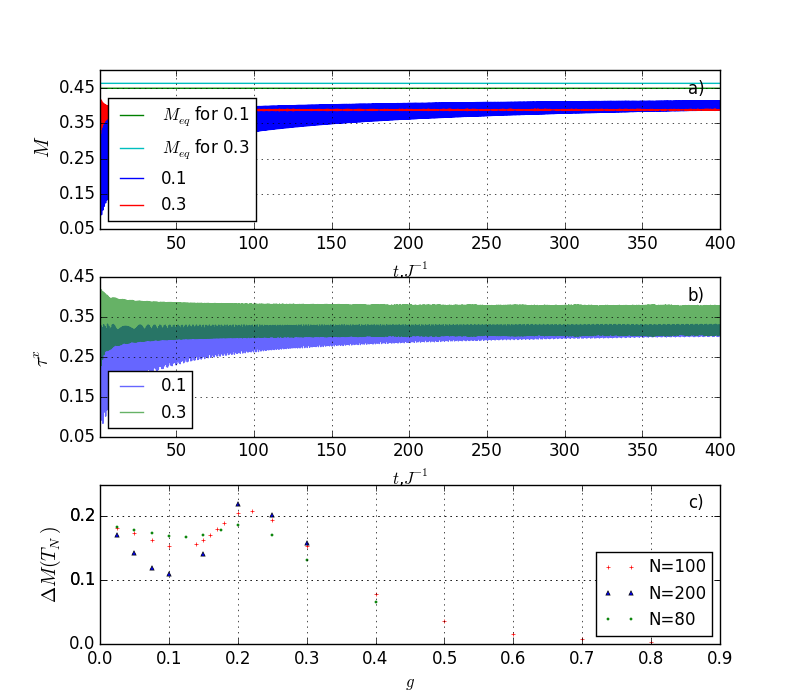}} 
	\setlength{\abovecaptionskip}{1pt plus 1pt minus 2pt}
	\caption{Impact of the bath on the post-quench dynamics. Equilibrated and damped regimes ($U_i=0.8$ to $U_f=12$, $\Lambda=20$). 
	a) Time evolution of staggered magnetization for several values of $g$ and $N=100$.
	b)  Time evolution of individual mode $\tau^x_k$ for several values of $g$ and $N=100$
	computed for a generic value of the momentum $k=\{ 2\pi/(N-1),\pi/(N-1) \}$. 
	c) Dimensionless parameter $\Delta M(T_N)$ as a function of $g$ for three values of $N$. The finite size times were taken to be $T_{N=80}=23$, $T_{N=100}=45$, $T_{N=200}=250$. 
	\label{fig:open:lockphase:panel}
	}
\end{figure}
For quenches with $U_f \gg U_i$,  roughly corresponding to the phase locked regime in a closed system, the presence of the dissipative bath leads to the decay of the persistent oscillations and the establishment of an asymptotic equilibrium state. 
Fig. \ref{fig:open:lockphase:panel}-a) shows the evolution of the order parameter for different values of the environment coupling $g$.
 For the smaller values of $g$ the long time  $M(t \to \infty)$ attains the equilibrium value. 
For the larger values of $g$ this is no longer the case. 
Fig. \ref{fig:open:lockphase:panel}-b) shows that these two asymptotic regimes correspond to the trivial $d \left< \hat{\boldsymbol \tau }_{\boldsymbol k }  \right> /dt= 0$ and non-trivial $d \left< \hat{\boldsymbol \tau }_{\boldsymbol k }  \right> /dt \neq 0$ dynamics of the pseudo-spins. 
Fig. \ref{fig:open:lockphase:panel}-c) shows the rescaled deviation from equilibrium of the order parameter $\Delta M(T_N) $ as a function of $g$ for different system sizes. The finite size scaling with $N$ shows that in the equilibrated phase $\Delta M(T_N)$ vanishes with increasing $N$ while for larger $g$ it attains a finite value. The presence of a fixed point around $g \approx 0.2$ indicates a dynamical phase transition between the two regimes.

The fact that the system does not equilibrate for large system-bath coupling seems rather counterintuitive. This can however be explained by the fact that besides dissipation, i.e. the appearance of a memory kernel in the evolution, the presence of the environment also renormalizes the coupling constant $U$ and thus the system moves into Landau-damping-like regime where equilibration with environment does not happen. 

The fact that the amplitude of oscillations that were persistent in closed system, now decreases with time can be understood in the following way: without bath, a phase-locked collective mode cannot transfer energy to individual modes due to the presence of the energy gap; in the open system there are always bath modes to which energy may be transferred.  As a result, excited state decays. Moreover, since the energy exchange between the collective and the individual quasiparticle modes is suppressed, there is no electronic relaxation mechanism available other then the bath. Therefore the only possibility for the system is to equilibrate with it.

\subsection{Overdamped regime}
\begin{figure}[htb!]
	\centerline{\includegraphics[width=0.5\textwidth]{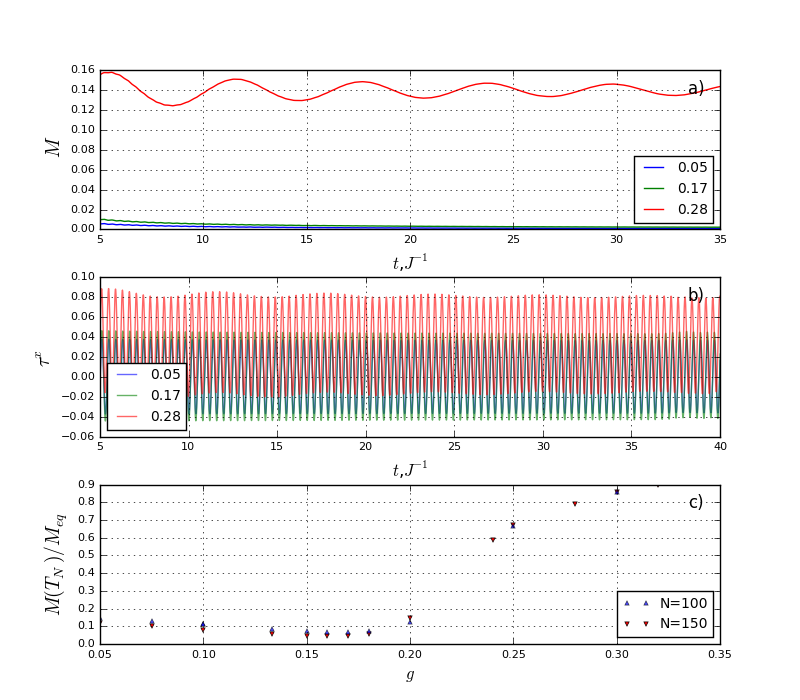}} 
	\setlength{\abovecaptionskip}{1pt plus 1pt minus 2pt}
	\caption{Impact of the bath on the post-quench dynamics. Overdamped and damped regimes ($U_i=3.0$ to $U_f=0.5$, $\Lambda=20$). 
	a) Time evolution of staggered magnetization for several values of $g$ and $N=150$.
	b)  Time evolution of individual mode $\tau^x_k$ for several values of $g$ and $N=150$
	computed for a generic value of the momentum $k=\{ 2\pi/(N-1),\pi/(N-1) \}$. 
	c) Dimensionless parameter $M(T_N)/M_{\text{eq}}$ as a function of $g$ for two values of $N$. The finite size times were taken to be $T_{N=100}=23$, $T_{N=150}=35$. 
	\label{fig:open:overdamped:panel}
	}
\end{figure}

In overdamped regime arising for $U_f \ll U_i$ the interaction with the bath does not change qualitatively the dynamics with respect to the $g=0$ case apart from the renormalization of $U$.
A fast decay of the order parameter to $M(\infty)=0$ can be observed for the small values of $g$ depicted in Fig. \ref{fig:open:overdamped:panel}-a).  For larger values of $g$ the damped regime sets in and $M(\infty)$ is non-vanishing. Fig.\ref{fig:open:overdamped:panel}-b) shows that even if $M(\infty)=0$, the microscopic dynamics $\left< \hat{ \boldsymbol{ \tau} }_{\boldsymbol k}  \right>  $  is non-trivial. 
In this regime the bath effectively decouples from the system since $M\approx 0$ and coupling to bath is proportional to $g^2M$.  As a consequence the system does not equilibrate and the order paramenter vanishes as a power law $M(t) \propto t^{\nu}$. 

\begin{figure}[htb!]
	\centerline{\includegraphics[width=0.4\textwidth]{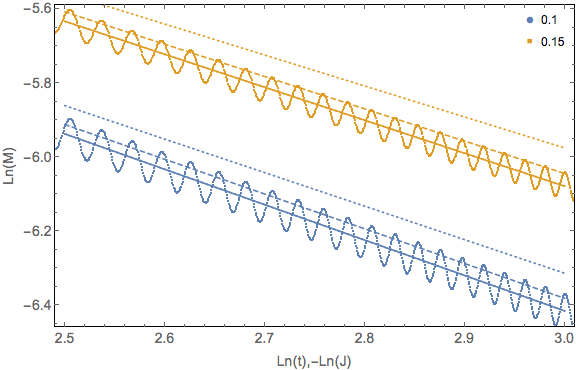}} 
	\setlength{\abovecaptionskip}{1pt plus 1pt minus 2pt}
	\caption{Exponents $\nu(g)$ in the overdamped ($U_i=3.0$ to $U_f=0.5$, $\Lambda=20$) regime for different values of coupling $g$ and different system sizes. Exponents were computed by fitting (solid line for linear size $N=200$, dashed for $N=150$, dotted for $N=100$) of the averages of time-dependence of logarithm of staggered magnetization (plotted only for $N=200$) by linear function $-\nu_N(g) \log(t)+a$.  
\label{fig:open:overdamped:exponent}
	}
\end{figure}

As in the damped regime the algebraic decay in the overdamped case is also $g$-dependent thus differing from the $\nu=1$ results obtained for a closed system. 
This is shown in Fig. \ref{fig:open:overdamped:exponent}, where  we have plotted the staggered magnetization averaged over a period in log-log scale. The numerical results are fitted to a linear function $\nu_N(g)y+a$, with $y=\log(t)$. $\nu$ is observed to vary with $g$:  $\nu_{200}(g=0.1)\approx 0.96$ to $\nu_{200}(g=0.18)\approx 0.70$. 
Notice that the result slightly varies with system size. This is because the equilibrium staggered magnetization is very sensitive to the size of the  system for these values of $U$. 
Nonetheless it is clear that  $\nu_{N}(g)$ is converging to a $g$ dependent exponent $\nu(g)$. 

The transition between the overdamped and the damped regimes upon increasing $g$ can be seen in the finite size scaling of $ M(T_N)/M(T_{\text{eq}})$ shown in Fig.  \ref{fig:open:overdamped:panel}-c). In the small $g$ region  $ M(T_N) $ vanishes for increasing $N$ whereas for large $g$ it seems to attain a finite value. 
The crossing of the finite size data is compatible with transition arising for $ 0.2 <g_c < 0.24 $ for the parameter values of Fig. \ref{fig:open:overdamped:panel}-c).

\section{Discussion}

We studied the dynamics ensuing after an interaction quench in a model consisting of a Hubbard layer coupled to an antiferromagnetic magnon bath. 
For vanishing system-bath coupling, within a mean-field approximation,  the post quench dynamics can be mapped to that of well studied BCS quenches. We identify the three known asymptotic long-time regimes: persistent oscillations of the order parameter, Landau-damping and overdamped. In the overdamped regime, we found that specificities of the 2d electronic density of states - a discontinuity at the band edges and a logarithmic divergence near the Fermi energy -  result in a power law decay of the order parameter rather than the exponential one reported for the BCS case that assumes a smooth density of states.  

For a finite system-bath coupling we show that the system does not always equilibrate as one would expect.
 Instead, three different regimes are observed at large times: 
 an equilibrating regime, where the system attains an asymptotic equilibrium state; 
 a damped regime, where the magnetization attains a static finite value that differs from the equilibrium one;  and an overdamped regime, characterised by an asymptotically vanishing magnetization.  

Each regime can be seen as a reminiscence of one of the different dynamical phases of the close system.
The persistent oscillations found in the phase-locked regime of the closed system do not survive in the presence of the bath and slowly decay to the equilibrium solution. 
The presence of non-equilibrium states is possible as the dissipative environment is only sensitive to time changes of the magnetization. For a static magnetization it acts simply as a renormalization of the Hubbard interaction. Therefore non-equilibrium phases with a static magnetization are stable. These static phases include zero magnetized phase of the overdamped regime and a phase similar to the one obtained in the Landau-damped regime of the closed system.  

We show that, though the bath does not change the dynamics qualitatively in the overdamped and damped regimes, there is a difference on how the staggered magnetization approaches its asymptotic value $M\simeq M(\infty)+O(t^{-\nu})$.
For finite $g$, the exponent $\nu$ does no longer take the discrete values $1$ or $1/2$. Instead, it seems to vary continuously in the range from $1/2$ to $1$ as a function of $g$.  

Our results are based on mean-field theory and therefore qualitatively correct only on time scales smaller than the quasiparticle lifetime $\tau_q$. The presence of the bosonic bath introduces an additional time-scale $\tau_g$. Thus our treatment is  relevant  for parameter sets such that $\tau_g\ll \tau_q$.
Moreover, mean-field approximation discard quantum fluctuations of the order parameter both perpendicular and parallel to the magnetization vector. It would be interesting to investigate the effect of these fluctuations in the asymptotic long time regime, in particular to study the survival of the non-equilibrium states found here at the mean-field level. 

Nonetheless, even if not all the dynamic mean-field regimes survive the inclusion of quantum fluctuation, traces of these regimes should be found at time scales for which fluctuation effects can be disregarded.


\acknowledgements 

We gratefully acknowledge discussions with Y.E. Shchadilova and enlightening remarks on the work of S.M. Apenko.
The study was founded by the RSF, grant 16-42-01057. PR acknowledges support by FCT through the Investigador FCT contract IF/00347/2014.

\appendix 

\section{Closed system - Crossover in overdamped regime} \label{Crossover}
Here we derive the asymptotic long time behaviour in the overdamped regime. 
Fig. (\ref{phase_diagram_close})-(lower-right panel) shows examples of the time evolution for $h_c(t=0)\ll8J$ and $ h_c(t=0)\gg8J$. 
Though the temporal behaviour for $h_c(t=0)\ll8J$ looks different from the one of $h_c(t=0)\gg8J$  both regimes can be described by a smooth function of $h_c(t=0)$.

In the overdamped regime which arises for $U_i/U_f\gg 1$, it is therefore natural to consider an expansion around small $U_f$.
At  $U_f=0$ one has that  $\boldsymbol B_{\boldsymbol{k}}(t>0) = \{0,0,\epsilon_{\boldsymbol{k}} \}$ and thus the evolution of the different momenta decouples.  Starting from initial conditions \eqref{initials} the evolution of the order parameter becomes 
\begin{equation}\label{analytics:M-integral}
M(t)
= \frac{1}{2} \int d\epsilon  \frac{\varrho(\epsilon)}{\sqrt{1+\frac{\epsilon^2}{h_c(t=0)^2}}} e^{2i\epsilon t}
\end{equation}
where $\varrho(\epsilon) = \int_k \delta( \epsilon-\epsilon_{\boldsymbol k})$ is the bare density of states of the electronic system. 

In two spatial dimensions $\varrho(\epsilon)$ has two distinctive features that may contribute to the asymptotic long time behaviour of $M(t)$:
 $\varrho(\epsilon) \propto -\ln(| \epsilon |)$ for $\epsilon\simeq 0$; and $\varrho(\epsilon)$  has sharp cutoffs at $\epsilon  = \pm 4J $. 
The damped-oscillatory or purely damped behaviour of $M(t)$ depends on the respective contribution of each of these features:
the denominator in the right-hand side of Eq.(\ref{analytics:M-integral})  defines a window of characteristic size $h_c(t=0)$ within which the integrand is non-negligible; if this window is much smaller than the bandwidth the only singularity that contributes to the long time behaviour is the one at $\epsilon\simeq 0$ that leads to an asymptotic behaviour in $1/t$; on the contrary, if the $h_c(t=0)$ is much larger than the bandwidth there are additional oscillatory contributions coming from the non-analyticities at the band edges that behave as $ \sin(8Jt)/t$. 

A more quantitative way to obtain the oscillatory-damped crossover of $M(t)$ as a function of $h_c(t=0)$ is to develop Eq.(\ref{analytics:M-integral}) around $h_c(t=0)\simeq4J$. Defining $\delta=h_c(t=0)- 4J$  we obtain: 
\begin{multline}\label{analytics:M-integral:delta-third-order}
M(t)
\approx
\frac{1}{8J\pi t}
+
\frac{\sin(8Jt)}{8J\pi\sqrt{2}t}
-
\frac{\delta}{2(4J)^2t\pi(2)^{3/2}}\sin(8Jt)
\\
-
\frac{3\delta^2}{4(4J)^3\pi 2^{5/2}t}
\sin(8Jt)
-
\frac{3\delta ^3}{4(4J)^4\pi 2^{7/2}t}
\sin(8Jt)
+
O(\delta^4)
\end{multline}
Even if the region of applicabшlity of this expression is limited, it reproduces well the numerical results and captures the crossover behaviour observed in the overdamped regime.

\bibliography{projectMFdynamics_v1}
\bibliographystyle{apsrev4-1}

\end{document}